\documentclass[twocol]{ametsocV6.1}

\usepackage[acronym]{glossaries}
\usepackage[colorlinks=true,linkcolor=blue,citecolor=blue]{hyperref}
\usepackage{layouts}
\usepackage{caption}
\usepackage{subcaption}
\usepackage{booktabs}
\usepackage{multirow}
\usepackage{makecell}
\usepackage{textcomp}
\usepackage{siunitx}
\usepackage[export]{adjustbox}
\usepackage{tabularx}
\usepackage{float}
\usepackage{balance}
\usepackage{float}

\renewcommand\vec[1]{\mathbf{#1}}

\newacronym{BQN}{BQN}{Bernstein Quantile Network}
\newacronym{DRN}{DRN}{Distribution Regression Network}
\newacronym{MOS}{MOS}{Model Output Statistics}
\newacronym{DNN}{DNN}{Deterministic Neural Network}
\newacronym{EMOS}{EMOS}{Ensemble Model Output Statistics}
\newacronym{MAE}{MAE}{Mean Absolute Error}
\newacronym{MLP}{MLP}{Multi-Layer Perceptron}
\newacronym{NN}{NN}{Neural Network}
\newacronym{NWP}{NWP}{Numerical Weather Prediction}
\newacronym{GDPS}{GDPS}{Global Deterministic Prediction System}
\newacronym{LBQ}{LBQ}{Linear Bernstein Quantiles}
\newacronym{CRPS}{CRPS}{Continuous Ranked Probability Score}
\newacronym{CRPSS}{CRPSS}{Continuous Ranked Probability Skill Score}
\newacronym{PIT}{PIT}{Probability Integral Transform}
\newacronym{LQR}{LQR}{Linear Quantile Regression}
\newacronym{QRN}{QRN}{Quantile Regression Network}
\newacronym{RMSE}{RMSE}{Root Mean Square Error}

\title{Leveraging deterministic weather forecasts for in-situ probabilistic temperature predictions via deep learning}

\authors{David Landry,\aff{a}\correspondingauthor{David Landry, david.landry@inria.fr} 
Anastase Charantonis,\aff{a,b,c,d} 
Claire Monteleoni\aff{a,e} 
}

\affiliation{
    \aff{a}{INRIA Paris}\\
    \aff{b}{ENSIIE, Evry }\\
    \aff{c}{LAMME, Evry }\\
    \aff{d}{LOCEAN/IPSL}\\
    \aff{e}{University of Colorado Boulder}
}

\abstract{
We propose a neural network approach to produce probabilistic weather forecasts from a deterministic numerical weather prediction.
Our approach is applied to operational surface temperature outputs from the Global Deterministic Prediction System up to ten-day lead times, targeting METAR observations in Canada and the United States.
We show how postprocessing performance is improved by training a single model for multiple lead times.
Multiple strategies to condition the network for the lead time are studied, including a supplementary predictor and an embedding.
The proposed model is evaluated for accuracy, spread, distribution calibration, and its behavior under extremes.
The neural network approach decreases CRPS by 15\% and has improved distribution calibration compared to a naive probabilistic model based on past forecast errors.
Our approach increases the value of a deterministic forecast by adding information about the uncertainty, without incurring the cost of simulating multiple trajectories. 
It applies to any gridded forecast including the recent machine learning-based weather prediction models. 
It requires no information regarding forecast spread and can be trained to generate probabilistic predictions from any deterministic forecast.
} 

\begin{document}

\maketitle

%
%
%
\statement
    Weather is difficult to predict a long time in advance because we cannot measure the state of the atmosphere precisely enough.
    Consequently, it is common practice to run forecasts several times and look at the differences to evaluate how uncertain the prediction is.
    This process of running ensemble forecasts is expensive and consequently not always feasible.
    We propose a middle ground where we add uncertainty information to forecasts that were run only once, using artificial intelligence.
    Our method increases the value of these forecasts by adding information about the uncertainty without incurring the cost of multiple full simulations.

\section{Introduction}

Weather forecast postprocessing is a key component in many operational forecasting systems.
When compared with in situ observations, \gls{NWP} models produce systematically biased forecasts, partly due to the presence of unresolved phenomena at finer scales.
This encourages the study of statistical postprocessing methods that correct these biases by training on past forecasting errors~\citep{VannitsemStatisticalPostprocessing2021}.
Earlier postprocessing models were used to generate probabilistic predictions from deterministic model outputs, notably for precipitation forecasting~\citep{AntolikOverviewNational2000}.
With the advent of ensemble forecasts, this approach has received less attention in recent years.
However, deterministic forecasts are still being produced in many operational centers~\citep{HamillComparingCombining2021}.
This is done at shorter lead times, for which deterministic forecasts are more appropriate.
It can also be motivated by computational constraints if different parts of the ensemble size/model resolution compromise are explored.

We propose a deterministic-to-probabilistic approach that recovers a deterministic model's uncertainty by training on past observations using \glspl{NN}.
Since it targets surface stations, it applies to any gridded \gls{NWP} including machine learning weather prediction models~\citep{LamGraphCastLearning2022,PathakFourCastNetGlobal2022,BiAccurateMediumrange2023}.
Our approach requires no input regarding forecast spread and infers all uncertainties from its training.

This contribution arises in a dynamic context for weather forecast postprocessing~\citep{VannitsemStatisticalPostprocessing2021}.
A topic of active discussion is how the distribution is predicted, i.e., the model that determines the shape of the output distribution.
Notable approaches include linear models~\citep{GneitingCalibratedProbabilistic2005}, random forests~\citep{TaillardatCalibratedEnsemble2016}, fully connected neural networks~\citep{RaspNeuralNetworks2018} and Convolutional Neural Networks~\citep{VeldkampStatisticalPostprocessing2021}.

Another topic is how the forecast uncertainty is represented, that is how the output of the predictive model is cast into a CDF.
Deterministic postprocessing is being studied at shorter lead times~\citep{HamillComparingCombining2021}.
For probabilistic outputs, solutions revolve around parametric~\citep{TaillardatSkewedMixture2021,DemaeyerEUPPBenchPostprocessing2023} and non-parametric~\citep{BremnesEnsemblePostprocessing2020,HewsonLowcostPostprocessing2021} methods (also called distribution-based and distribution-free methods, respectively).
Among the non-parametric methods, \citet{BremnesEnsemblePostprocessing2020} proposes a \gls{BQN} that models the quantile function of the probabilistic forecast as a Bernstein polynomial.
This approach yields improvements over previous methods for surface wind speed fields while avoiding strong assumptions about the distribution of the predicted value.
Numerous parametric and non-parametric methods were systematically reviewed for wind gust forecasting by \citet{SchulzMachineLearning2022}, using a modular framework that decouples how the distribution is predicted (i.e. using a linear model or a neural network) from how it is represented (i.e. using a normal distribution or quantile regression).

Finally, recent works have trained postprocessing models separately for all lead times~\citep{RaspNeuralNetworks2018,BremnesEnsemblePostprocessing2020,SchulzMachineLearning2022} or jointly in a single model~\citep{BouallegueStatisticalModeling2023}.
In the latter, it was observed that the lead time predictor was not always retained by a feature selection process for deterministic forecast postprocessing.
As such, there are discussions regarding how to best condition a postprocessing model for the lead time.

The deterministic-to-probabilistic approach has seen renewed interest in recent work.
\citet{VeldkampStatisticalPostprocessing2021} proposed a convolutional network to postprocess wind speed forecasts from a high-resolution deterministic model.
Their prediction is done at a single lead time.
\citet{BouallegueStatisticalModeling2023} proposed a two-step postprocessing pipeline where a second model is trained to predict the residual error of the first model.
The residual error model provides an estimation of the uncertainty of a deterministic forecast but does not express its probability distribution.
\citet{BremnesEvaluationForecasts2023} applied a \gls{BQN} to deterministic weather forecasts produced by deep neural networks~\citep{BiAccurateMediumrange2023}, making a probabilistic forecast from a deterministic \gls{NWP}. 
Their work studies only a quantile-based method and invites the evaluation of other approaches.
\citet{DemaeyerEUPPBenchPostprocessing2023} introduced ANET which predicts a parametric forecast using a single network for all stations and lead times. 
It is compatible with variable member counts.

We perform further examinations by producing probabilistic forecasts based on a deterministic global model, with up to ten-day lead times.
This is done on a dataset built from the \gls{GDPS} \gls{NWP} model~\citep{BuehnerImplementationDeterministic2015}, targeting METAR surface temperature observations in Canada and the United States~\citep{METARDataset}.
We study neural network models for this purpose and compare them to linear models. 
The forecast uncertainty is expressed using either a normal distribution, a set of quantile values, or a quantile function built from Bernstein polynomials.
The neural network models are trained jointly for all lead times.
We compare multiple strategies to condition for it, including a lead time embedding that has shown usefulness in recent work~\citep{EspeholtDeepLearning2022}. 
Finally, we evaluate the behavior of our post-processing method under extreme temperatures and compare its behavior to that of the \gls{NWP} model.

The next section describes our experimental framework in more detail, including models, datasets, and evaluation methods.
Section \ref{sec:experiments} contains our experiments that evaluate forecast performance, calibration, as well as its behavior under extreme events.
This is followed by a discussion about the proposed postprocessing methods and our concluding remarks in Section \ref{sec:discussion}.

\section{Methods}
\label{sec:methods}

\subsection{Data}
\label{ssec:data}

We perform postprocessing of surface temperature fields over the operational output of the \gls{GDPS} NWP model~\citep{BuehnerImplementationDeterministic2015,GDPSDataset} and target observations from the METAR network.
We use model outputs initialized at 0000 and 1200 UTC to perform postprocessing daily up to 10-day lead times.

For training, we use forecasts initialized from 1 January 2019 to 31 December 2020.
This period contains a major update of the GDPS model in July 2019, where the model horizontal resolution was increased from 25 to 15 km~\citep{McTaggart-CowanModernizationAtmospheric2019}.
We still retain the earlier forecasts, since removing them provoked a slight decrease in validation score.

Forecasts initialized on the first 25 days of each month are used for the training itself, while the others are used for validation.
This validation strategy is not fully independent because late lead times from the training set overlap with early lead times from the validation set.
We choose this compromise to ensure good coverage of the seasonal cycle.
Forecasts from 1 January 2021 to 30 November 2021 are used for testing.
We note that our testing period for the GDPS dataset contains two extreme weather phenomena, a wave of extreme cold temperatures in Texas in February, and a heat wave over western North America in June and July.

We use 18 \gls{NWP}-dependent predictors from the GDPS dataset, as well as 7 NWP-independent predictors.
Table \ref{tab:features} contains the full predictor list.
The set of NWP-dependent predictors is relatively small due to constraints in accessing and storing operational outputs.

All features are scaled using their mean and variance over the training set to improve normality.
This is done station-wise.
Before scaling, we apply a logarithm transformation on positively defined variables (albedo, precipitation, and wind speed).
We provide the day-of-year predictor twice, encoded with sin and cos transforms, to represent periodicity.

\begin{table*}

    \centering
    \caption{NWP-dependent and NWP-independent predictors used for postprocessing. 
    \checkmark = Always used; * = used unless specified otherwise.}

    \small
    \begin{tabular}{llc}
    \toprule
    Type & Predictor  & Vertical level/Usage \\
    \midrule
    NWP-Dependent &Albedo        & Surface \\
    &Dew point temperature       & 2 m \\
    &Geopotential Height         & \{1000,850,500\} hPa \\
    &Mean sea-level pressure     & Sea level datum\\
    &Precipitation rate          & Surface \\
    &Relative humidity           & 2 m \\
    &Specific humidity           & \{850,500\} hPa \\
    &Temperature                 & 2 m, \{850,500\} hPa \\
    &{U,V} component of wind     & 10m, 500 hPa \\
    &Wind speed                  & 10 m \\
    \midrule
    NWP-independent&Lead time & *\\
    &Forecast day-of-year (sin and cos)  & \checkmark\\
    &Forecast time of day                & \checkmark \\
    &Latitude                            & \checkmark\\
    &Longitude                           & \checkmark\\
    &Elevation                           & \checkmark\\
    \bottomrule
    \end{tabular}\label{tab:features}
    \label{tab:features}
\end{table*}

We target observations from the METAR network harvested from the Iowa State University Environmental Mesonet~\citep{METARDataset}.
The dataset consists in observations from 1066 stations spread across North America. 
The \gls{NWP} forecasts were interpolated to stations using the nearest grid point.
Stations were selected based on data availability for the periods covered by the \gls{NWP} forecasts.
We removed 250 observations ($\sim$0.01\%) that reported temperatures more than 15 K off their corresponding 24h forecast after debiasing.
The bias values were computed by averaging model errors over the training set, separately for each station, initialization time, lead time, and month.
This process is meant to account for sensor errors.

\subsection{Postprocessing Models}
\label{ssec:models}

We devise a naive probabilistic forecast as a baseline model.
First, the \gls{NWP} forecasts are debiased using the process described in Section \ref{ssec:data}.
Secondly, the standard deviation of forecast errors is computed using the same aggregation (separately by station, initialization time, lead time, and month).
Our baseline probabilistic forecast is a normal distribution centered around the debiased \gls{NWP} forecast and scaled by the computed standard deviation.

Our other methods use machine learning to generate probabilistic forecasts from \gls{NWP}.
They are summarized in Table \ref{tab:all_labels}.
We follow a decoupled approach~\citep{SchulzMachineLearning2022} where we distinguish how the forecast uncertainty is predicted from how it is represented.
We study two predictive models: a linear model and a \gls{NN} model.
We study four uncertainty representations: a naive deterministic representation, a normal distribution, a set of quantiles, and a quantile function posed as a Bernstein polynomial.

\begin{figure*}
    \centering
    \includegraphics{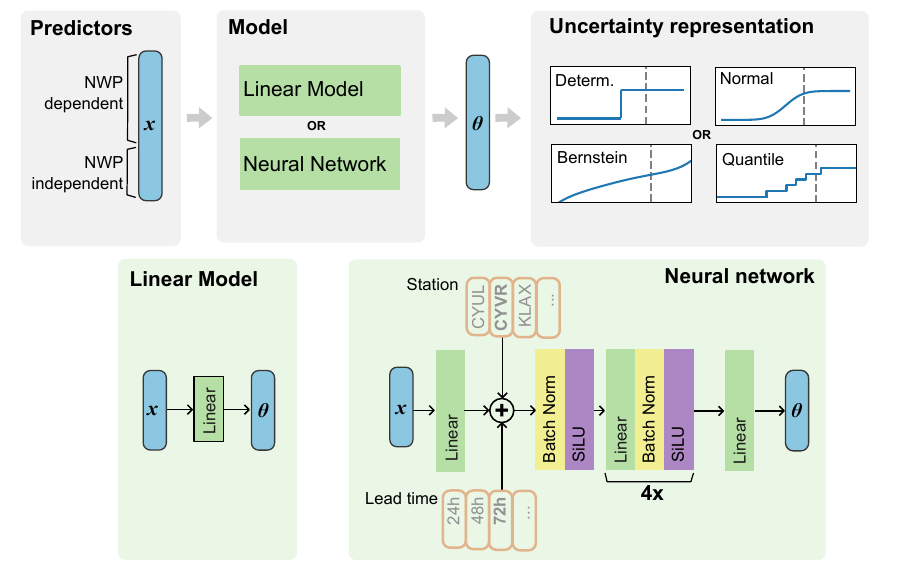}
    \caption{
        \textbf{Top}. Our model architecture. 
        Each forecast is made using one of two parameter prediction models, and one of four uncertainty representations.
        \textbf{Bottom left}. The linear model predicts a parameter vector $\boldsymbol{\theta}$ using one linear layer. 
        \textbf{Bottom right}. The \gls{NN} predicts a parameter vector $\boldsymbol{\theta}$ by processing the predictors through a \acrfull{MLP}. 
        It has embeddings for station and lead time, which are added to the output of the first linear layer.
    }   \label{fig:architecture}
\end{figure*}

This yields eight different models which are illustrated in Figure \ref{fig:architecture}.
The models use NWP-dependent and NWP-independent predictive features $\boldsymbol{x}$.
These features are used by a predictive model to build a vector $\boldsymbol{\theta}$ which determines the predictive distribution. 
One such vector is produced for each station and lead time.
It defines the probabilistic forecast, for instance by providing its distribution parameters or a set of quantile values.
The length of $\boldsymbol{\theta}$ changes according to which uncertainty representation is used.

\begin{table*}
    \small
    \centering
    \caption{Summary table of the postprocessing models considered in this work.}

    \begin{tabular}{cccc}
    \toprule
         &  \multicolumn{3}{c}{Predictive model}\\
         \makecell{Uncertainty\\representation}   & Linear & Neural network & \makecell{Length of $\boldsymbol{\theta}$} \\
    \midrule
         Deterministic          & \makecell{\acrlong{MOS} \\ \acrshort{MOS} \citep{GlahnUseModel1972}}  & \makecell{Deterministic Neural Network\\DNN \citep{BouallegueStatisticalModeling2023}} & 1  \\[3.5ex]
         Normal distribution            & \makecell{\acrlong{EMOS} \\ \acrshort{EMOS} \citep{GneitingCalibratedProbabilistic2005}}  & \makecell{\acrlong{DRN} \\ \acrshort{DRN} \citep{RaspNeuralNetworks2018}} & 4 \\[3.5ex]
         Bernstein polynomial   & \makecell{\acrlong{LBQ} \\ \acrshort{LBQ}} & \makecell{\acrlong{BQN} \\ \acrshort{BQN} \citep{BremnesEnsemblePostprocessing2020}} & 16\\[3.5ex]
         Quantiles              & \makecell{Linear Quantile Regression \\ LQR \citep{BremnesProbabilisticForecasts2004}} & \makecell{\acrlong{QRN} \\ \acrshort{QRN} \citep{CannonNoncrossingNonlinear2018}} & 32 \\
    \bottomrule
    \end{tabular}
    \label{tab:all_labels}
\end{table*}

\subsubsection*{Predictive models}

Our first parameter prediction model linearly maps the predictors $\boldsymbol{x}$ to $\boldsymbol{\theta}$.
It has separate coefficients and biases for each station, initialization hour, and lead time.

Our second model is a \gls{NN} model centered around a \gls{MLP}, as introduced for postprocessing by \citet{RaspNeuralNetworks2018}.
The hidden layer size is kept constant throughout.
The SiLU activation function~\citep{RamachandranSearchingActivation2017} was used.
Batch norm layers were added after the linear layers since we observed they accelerate convergence during training.

It is common to train the NN model jointly for all stations and then condition for the station with an embedding~\citep{RaspNeuralNetworks2018}.
An embedded station vector is learned for each station.
It has the same size as the hidden layers in the \gls{MLP}.
It is added to the network activations after the first linear layer, but before the batch norm and activation layers are applied.
Doing the addition in that location is equivalent to concatenating one-hot encoded predictors to $\boldsymbol{x}$ representing the station identity.
Our embedding approach has a smaller memory footprint because it avoids padding $\boldsymbol{x}$ with numerous and sparsely populated features, which facilitates implementation.

Since our model is trained for 10 lead times on the GDPS dataset, we need to condition for the lead time as well.
We use two strategies: we add a predictor to $\boldsymbol{x}$ corresponding to the lead time (rescaled from 0 to 1).
We also learn a lead time embedding implemented in the same way as the station embedding.
This lets the network adapt its postprocessing strategy to the lead time (e.g. by increasing uncertainty), but also identify correlation structures that are common across lead times.
The lead time embedding was already proposed in precipitation forecasting~\citep{EspeholtDeepLearning2022}.

\subsubsection*{Uncertainty representation}
\label{ssec:uncertainty}

Four ways to represent the postprocessed forecast uncertainty are considered.
Each uncertainty representation method accepts a  parameter vector $\boldsymbol{\theta}$ from one of the predictive models.
Each method has a corresponding loss function so that the models can be trained against observations.

In the base case, we consider a one-parameter output and use it as a deterministic forecast.
When the underlying predictive model is linear, we refer to this method as \gls{MOS}~\citep{GlahnUseModel1972}.
When it is a neural network, we refer to it as a \gls{DNN}.
These models are trained using the \gls{RMSE} loss function.

Our second approach follows from \gls{EMOS}~\citep{GneitingCalibratedProbabilistic2005} and \glspl{DRN}~\citep{RaspNeuralNetworks2018}.
Here, $\boldsymbol{\theta} \in \mathbb{R}^4$ such that the postprocessed forecast $Y_{obs}$ is 
\begin{align}
Y_{obs} &\sim \mathcal{N}(\theta_1 x_{nwp}  + \theta_2, \exp (\theta_3 \log (\hat{\sigma}) + \theta_4))
\end{align}
where $x_{nwp}$ is the raw \gls{NWP} forecast for surface temperature and $\hat{\sigma}$ is an initial guess for the forecast distribution standard deviation.
The initial guess $\hat{\sigma}$ would typically be the ensemble standard deviation if an ensemble forecast was being postprocessed.
In the absence of an ensemble forecast, it is substituted with the standard deviation of forecast errors as defined for the naive probabilistic model (aggregated by station, initialization time, lead time, and month).
We use the exponent and logarithm on the scale parameter to preserve positiveness.
We fit this model using the \gls{CRPS} as a loss function.
A derivation of the CRPS for normal distributions is provided in subsection \ref{ssec:evaluation}.

The quantile regression approach consists in predicting $\boldsymbol{\theta} \in \mathbb{R}^n$ containing the $n$ values of quantiles at positions $\tau \in \{\frac{1}{n+1}, \frac{2}{n+1}, ..., \frac{n}{n+1}\}$.
We refer to this method as \gls{LQR} in its linear variant~\citep{BremnesProbabilisticForecasts2004} and \gls{QRN} in its \gls{NN} variant.
We fit the quantiles using the quantile loss defined in Section~\ref{ssec:evaluation}.
Neural quantile regression has been approached with architectures that enforce quantile monotonicity by construction \citep{CannonNoncrossingNonlinear2018}.
We adopt an empirical approach where the quantiles are reordered after each prediction.
Note that the quantile locations described here are not \gls{CRPS}-optimal.
They were chosen to train models that target a uniform rank histogram~\citep{BrockerEvaluatingRaw2012} to simplify interpretation.

The Bernstein polynomial approach consists in learning a Bernstein polynomial that represents the forecast distribution quantile function~\citep{BremnesEnsemblePostprocessing2020}.
We call these approaches \gls{LBQ} and \acrfull{BQN} when the coefficients are predicted by a linear model and a \gls{NN}, respectively.
The quantile function $Q(\tau)$ is parameterized by a j-vector $\boldsymbol{\theta}$ such that
\begin{align}
    Q(\tau) &= \sum_{j=0}^d \theta_j \binom{d}{j} \tau^j (1 - \tau)^{d-j},
\end{align}
where $d$ is the degree of the Bernstein polynomial and $\binom{d}{j}$ is the binomial coefficient.
One such polynomial is produced for each forecast.
To compute loss values from $Q(\tau)$, we sample it at $n=98$ evenly spread locations such that $\tau \in \{\frac{1}{n+1}, \frac{2}{n+1}, ..., \frac{n}{n+1}\}$.
Then, the sampled values are evaluated using the quantile loss.
In contrast with our \gls{LQR} and \gls{QRN} models, $\boldsymbol{\theta}$ does not contain quantile values directly, but rather the coefficients that shape the quantile function.

Bernstein polynomials have shape-preserving properties where a monotonous list of coefficients $\theta_j$ will yield a monotonous function $Q(\tau)$, an expected property for quantile functions.
We enforce this by sorting the coefficients $\theta_j$ after they are predicted.
We observed that without coefficient ordering, the \gls{BQN} would sometimes converge to solutions that have good validation scores but poor calibration due to creating jagged quantile functions.

\subsubsection*{Combining predictions}

\gls{NN} methods converge to different models depending on their random weight initialization and the random composition of training batches.
To account for this, \gls{NN}-based postprocessing models are typically trained multiple times and their predictions are combined to create the final distribution~\citep{SchulzAggregatingDistribution2022}.
We follow this trend and train each \gls{NN} five times.
To combine predictions, we average the parameter vectors $\boldsymbol{\theta}$ across models.
For the \gls{DRN}, this is equivalent to averaging distribution parameters as is done by \citet{RaspNeuralNetworks2018}.
For the \gls{QRN}, this is equivalent to uniform weight quantile averaging, also known as Vincentization~\citep{SchulzAggregatingDistribution2022}.
This is also equivalent for the \gls{BQN} despite the fact that the values of $\boldsymbol{\theta}$ represent polynomial coefficients instead of quantile values~\citep{SchulzMachineLearning2022}.

\subsection{Training}

All models are implemented in PyTorch~\citep{PytorchSoftware}, a scientific computation framework well suited to deep learning applications.
They are trained using the Adam optimizer~\citep{KingmaAdamMethod2014}.
They are trained for 100 epochs with the OneCycleLR training scheduler~\citep{SmithSuperConvergenceVery2018}.
The maximal learning rate was $10^{-3}$ for the linear model and $5 \times 10^{-4}$ for the \gls{NN}.
A weight decay of $10^{-5}$ was used in both instances.

Each model was trained in four variants, corresponding to each of the uncertainty representations described in section \ref{ssec:models}.
We optimize variant-specific details for each model via a manual grid search (i.e. degree of Bernstein polynomial and number of predicted quantiles).
The shape of the shared architecture (embedding size and number of fully connected layers) was kept fixed across the variants to facilitate model intercomparison.

\subsection{Evaluation}
\label{ssec:evaluation}

To evaluate the quality of our probabilistic forecasts, we first use the \gls{CRPS} metric~\citep{GneitingStrictlyProper2007}.
For \gls{EMOS} and \gls{DRN}, we relate an observation $y$ to the predicted normal distribution with mean $\mu$ and standard deviation $\sigma$ using its closed-form expression~\citep{GneitingCalibratedProbabilistic2005} where
\begin{multline}
    CRPS_{norm} = \sigma \left[ \frac{y-\mu}{\sigma} \left(2 F \left(\frac{y-\mu}{\sigma} \right) - 1\right) +  \right. \\
    \left. 2f\left(\frac{y-\mu}{\sigma}\right) - \frac{1}{\sqrt{\pi}} \right].
\end{multline}
Here, $F$ and $f$ are the CDF and the PDF of the normal distribution, respectively.

For quantile-based forecasts, we interpret the quantile values as an ensemble forecast. 
In that situation, we compute the \gls{CRPS} using
\begin{align}
    CRPS_{ens} = \frac{1}{n} \sum_{i=1}^n | e_i - y |  - \frac{1}{2n^2}\sum_{i=1}^n\sum_{j=1}^n | e_i - e_j |
    \label{eqn:crps-ensemble}
\end{align}
for an $n$ member ensemble forecast $\{e_1...e_n\}$. 
Our work compares the performance of distributions for which we have closed-form CRPS expressions (\gls{EMOS}, \gls{DRN}) against distributions where it is approximated numerically (\gls{LBQ}, \gls{BQN}).
Care must be taken in the choice of \gls{CRPS} estimator because it should be made according to the properties of the ensemble forecast.
Equation \ref{eqn:crps-ensemble} is a better option here because the quantile values produced by our models are not exchangeable~\citep{ZamoEstimationContinuous2018}. 
However, it is known to have biases for small numbers of quantiles, which should be kept in mind when analysing our results.
The CRPS can also be evaluated with respect to a reference method, in which case it becomes the \gls{CRPSS}, defined as 
\begin{align}
    CRPSS &= 1 - \frac{CRPS_{model}}{CRPS_{baseline}}.
\end{align}

We measure forecast accuracy at the tails using the quantile loss $QL_\tau$, defined as 
\[
    QL_\tau = \left\{ \begin{array}{lr}
        \tau(y - q_\tau), & q_\tau \leq y       \\
        (1 - \tau) (q_\tau - y), & q_\tau > y 
    \end{array}\right.
\]
where $\tau$ is the quantile level being evaluated, $q_\tau$ is the corresponding quantile value, and $y$ the observation.
For normal distribution forecasts, $q_\tau$ is computed exactly using the inverse CDF.
For quantile-based approaches, we use a linear interpolation between the two nearest available quantiles.

To measure forecast uncertainty, use the composite spread as proposed by~\citet{BremnesConstrainedQuantile2019}.
For quantile forecasts, the smallest quantile interval widths are summed until they cover the desired probability range (here 80\%).
A linear interpolation is used in the final interval to reach our desired probability interval more precisely.
For normal distribution forecasts, this reduces to computing the distance between the 10th and 90th percentiles, as computed using the inverse CDF.
Finally, we assess the calibration of the predicted distributions using rank histograms.

\section{Experiments and Results}
\label{sec:experiments}

\subsection{Hyperparameter optimization}

For the \gls{NN} models, we choose four hidden layers with a size of 256, after observing diminishing returns on the validation set for larger architectures.
We did not find strong interactions between shared parameters and parameters related to the uncertainty representation, i.e. the optimal embedding size performed similarly on the \gls{DRN}, \gls{QRN}, and \gls{BQN}.
Regarding the length of the $\boldsymbol{\theta}$ vectors, we use Bernstein polynomials of degree 16 after observing no performance improvement for larger values.
We use 32 quantiles in our quantile regression models for the same reason.
The values considered were $\{8,10,12,14,16,18\}$ for the degree of the Bernstein polynomial and $\{16,20,...,40\}$ for the number of predicted quantiles of the quantile regression model.
The linear quantile methods (\gls{LQR}, \gls{LBQ}) were set to use the same length of $\boldsymbol{\theta}$ as their \gls{NN} counterpart to compare them at equivalent forecast resolution.

Once these hyperparameters are set, our NN models have between 580k and 590k trainable parameters.
The variability is due to the different number of output parameters $\boldsymbol{\theta}$ for each model.
For comparison, the EMOS model has 1,517,164 trained parameters.
It has more trained parameters than the NN because it has separate coefficients for each station, initialization time, and lead time.

\subsection{Postprocessing performance}
\label{ssec:pp-performance}

We ran a series of experiments to evaluate the performance of our postprocessing methods over the GDPS dataset.
Table \ref{tab:scores} shows performance metrics aggregated for all lead times and all stations.
The \gls{NN} models score better than their linear model in all configurations.
The linear model performs better with the normal distribution, while the \gls{NN} model shows similar performance for all variants.
We posit this is because the linear model does not have enough representation capability to correctly predict a large amount of interrelated $\mathbf{\theta}$ parameters.
Among \gls{NN} models, the \gls{BQN} obtains slightly better results for all evaluated metrics.

\begin{table*}
    \centering
    \small
    \caption{
    Postprocessing model metrics, aggregated over all stations and lead times.
    The \gls{CRPSS} is computed against the naive probabilistic baseline.}

\begin{tabular}{lcccccccccc}
    \toprule
    &\multicolumn{2}{c}{NWP}  & \multicolumn{4}{c}{Linear}  & \multicolumn{4}{c}{Neural Network} \\
    \cmidrule(lr){2-3} \cmidrule(lr){4-7} \cmidrule(lr){8-11}
    Model       &Raw    & Naive  & MOS   & EMOS            & LBQ & LQR     & DNN      & DRN & BQN & QRN \\
    \midrule
    CRPS        & 2.925	& 1.921	& 2.467	& \textbf{1.700}	& 1.852	& 1.897	& 2.315  & 	1.633 & \textbf{1.622} & 1.635 \\
    CRPSS       & -0.523	& 0.000	& -0.284	& \textbf{0.115}	& 0.036	& 0.012	& -0.205  & 0.150 & \textbf{0.156} & 	0.149 \\
    RMSE        & 4.070	& 3.783	& 3.385	& \textbf{3.334}	& 3.566	& 3.597	& 3.256	& 3.227	& \textbf{3.216}	& 3.227 \\
    QL$_{0.05}$  & - 	& 0.392     & - 	& \textbf{0.346}           &	0.427 &	0.462   &	-     &	0.335 &\textbf{	0.320}	& 0.323 \\
    QL$_{0.95}$   & -     & 0.375	    & - 	& \textbf{0.327}            & 0.402 &	0.436    &	- &	0.301 & \textbf{0.291}	& 0.295\\
    \bottomrule
\end{tabular}

\label{tab:scores}
\end{table*}

\begin{figure*}
    \centering
    \includegraphics{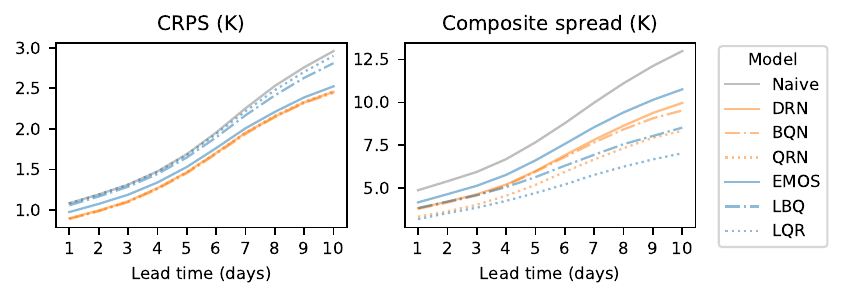}
    \caption{
    Postprocessing model metrics.
    \textbf{Left}: CRPS by lead time.
    \textbf{Right}: Forecast sharpness as measured by a composite spread covering 80\% of the predicted forecast distribution.}
    \label{fig:lead-time-metrics}
\end{figure*}

Figure \ref{fig:lead-time-metrics} shows metrics describing forecast behavior across lead times.
The CRPS and spread increase with lead time, as expected.
Forecasts from the naive probabilistic model have more spread throughout. 
The \gls{EMOS} model is successful in increasing sharpness, but less than the \gls{NN} models.
The \gls{QRN} has the lowest spread among \gls{NN} models, although this could be due to bias in the spread estimation related to its smaller number of quantiles.

\begin{figure*}
    \centering
    \includegraphics{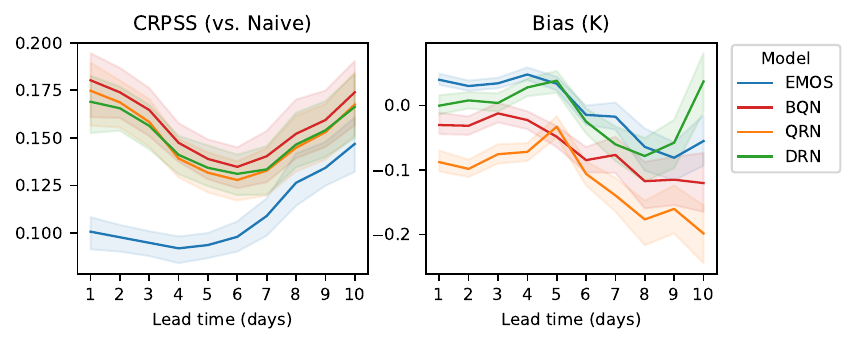}
    \caption{
    \textbf{Left:} \gls{CRPSS} against the naive probabilistic model. 
    \textbf{Right:} Bias of postprocessing models. 
    The shaded areas represent a 5 to 95\% confidence interval.}
    \label{fig:crpss-bias}
\end{figure*}

Model \gls{CRPSS} and bias are reported in Figure \ref{fig:crpss-bias}.
The confidence intervals are computed using the paired bootstrapping procedure proposed by 
\citet{HamillHypothesisTests1999}, under the null hypothesis that the naive probabilistic model has the same statistic as the tested model.
The resampling was performed 100 times.
The \gls{CRPSS} gains brought by \gls{NN} models are significantly larger than those of \gls{EMOS} at early lead times.
This difference reduces to 2.5\% for later time steps.
Bias values are computed by comparing the observation to the mean quantile value for models that output quantiles, while they are compared against the forecast mean for the \gls{EMOS} and \gls{DRN} models.
The \gls{NN} models do not eliminate all biases at longer lead times, though they remain under 0.3 K of amplitude.

\begin{figure}
    \centering
    \includegraphics[width=\columnwidth]{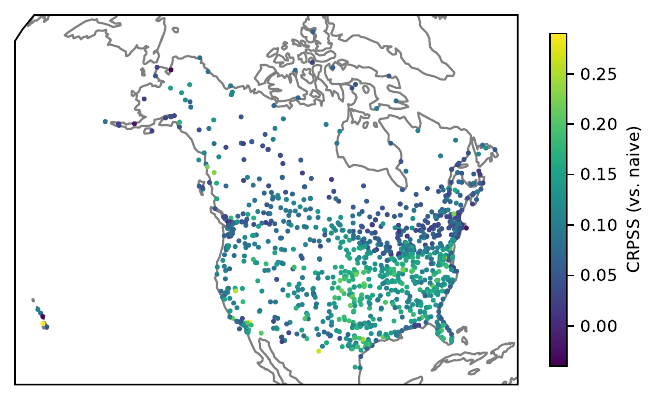}
    \caption{\gls{CRPSS} aggregated by station for all lead times.
    The model is \gls{DRN}.
    The baseline model is the naive model probabilistic model.}
    \label{fig:skill-spatial}
\end{figure}

Figure \ref{fig:skill-spatial} shows the \gls{CRPSS} against the naive probabilistic model, aggregated by station.
Larger increases in \acrshort{CRPSS} are observed in central and eastern United States.
We posit this reflects the location of some surface temperature biases in the underlying \gls{NWP} model.
The figure shows results for the \gls{DRN}.
Similar figures for the \gls{BQN} and \gls{QRN} are available in the appendix.

\begin{figure*}
    \centering
    \includegraphics[]{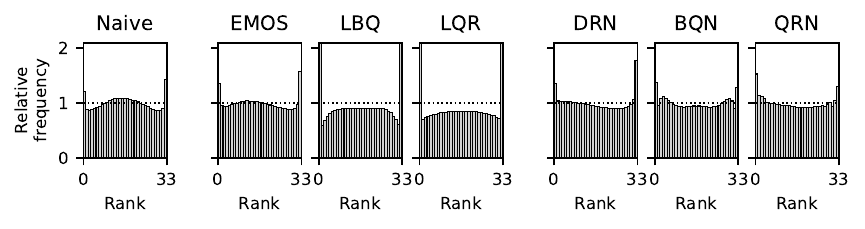}
    \caption{
    Postprocessed forecast rank histogram.
    For the normal uncertainty representation models (naive, \gls{EMOS}, \gls{DRN}), bin boundaries with uniform probabilities were computed using the inverse CDF of the forecasts.
    For the \gls{LBQ} and \gls{BQN} models, bins were merged in groups of three to allow comparison with the other histograms.}
    \label{fig:rank-histogram}
\end{figure*}

The forecast calibration is assessed using the rank histograms in Figure \ref{fig:rank-histogram}.
For the \gls{LBQ} and \gls{BQN} models, the 99 bins were merged in groups of three so that the rank histogram would be similar to that of the quantile methods.
For the methods producing normal distributions (naive, \gls{EMOS}, \gls{DRN}), their inverse CDF was discretized in 33 bins for the same reason.

All models have a surplus of observations at the first and last bins, indicating a certain amount of them are at the very edges of the forecasted distribution.
Other than these missed forecasts, the postprocessing models tend to flatten the central part of the rank histogram when compared to the naive model.
The \gls{BQN} shows interesting shapes at the edges.
We postulate they reflect what distributions can be expressed using a 16th-degree Bernstein polynomial.
Even though our validation scores stopped improving for higher-degree polynomials, they may be worth investigating from a calibration perspective.

\subsection*{Behavior towards extremes}

\begin{figure*}
    \centering
    \includegraphics[]{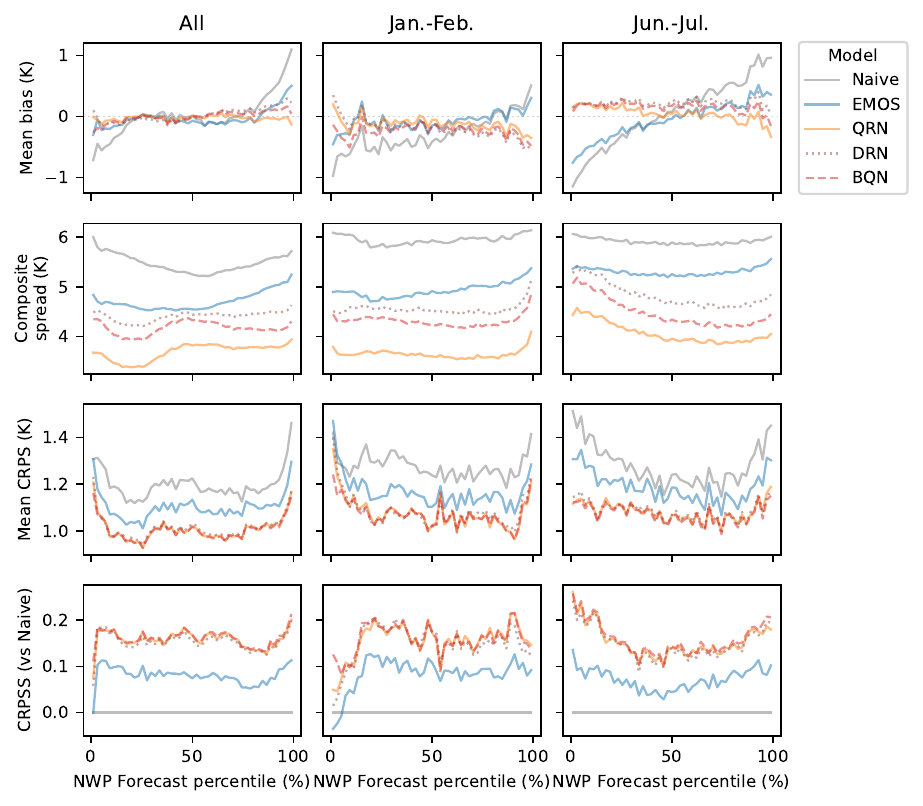}
    \caption{Model metrics according to the NWP forecast percentile.
    Lead time 48h.
    Initialization time 0000 UTC.
    The percentiles are computed station-wise.
    The forecasts are aggregated in bins of 2 percentiles.
    The metrics are computed over three periods: full test set (All months, left column), winter (January and February, middle column) and summer (June and July, right column).}
    \label{fig:extremes-by-forecast}
\end{figure*}

We evaluate the behavior of the NN models when \gls{NWP} forecasts tend towards extreme lows and highs in Figure \ref{fig:extremes-by-forecast}. 
Predictions were aggregated according to the percentile of their corresponding forecast for a given station.
They are grouped in bins of two percentiles.
The metrics are computed for an initialization time 0000 UTC and a lead time of 48 hours.

The naive model has biases when the forecast percentile tends to the extremes.
This is an expected consequence of our forecast-based stratification~\citep{BellierSampleStratification2017}.
Although the yearly curves for CRPS are flat in the central quantiles, more specific evaluations reveal that the all models have degraded predictive performance given very high and low forecasts within a season.
We observe a particularly strong degradation in \gls{CRPSS} for low percentiles in winter.
Upon inspection, these bins are disproportionately populated with forecasts valid from 12 February 2021 to 17 February 2021.
These dates are associated with unusually cold weather events in Texas which are outside the training distribution.

\subsection*{Conditioning for lead time}

Our \gls{NN} models are trained jointly for all lead times, which implies it needs to be conditioned for the lead time being predicted.
We introduced a lead time embedding in section \ref{sec:methods} for that purpose, as well as the typical lead time predictor.
This section studies the effectiveness of these strategies and analyzes the representation learned by the embedding.

In Table \ref{tab:condition}, we compare the CRPS values obtained by each strategy on the full testing set.
As baselines, we include results for non-conditioned models, as well as another strategy called Partitioning.
It consists in training separate models for each lead time.
We call this strategy partitioning because it effectively splits the dataset into parts.
We do not test the embedding on the linear model, because it uses separate models for each lead time.

In all cases, training the postprocessing model jointly for all lead times performed better than training separate models, given that the joint model is conditioned for lead time in one way or the other.
The lead time embedding improves performance slightly across all models when compared to using only a predictor.
The best results were obtained by using it together with the lead time predictor, except for the \gls{QRN} where the embedding alone had best performance.
Our dataset did not include diurnal variations: each lead time points to the same time of day.
Models trained for multiple lead times per day may see more benefits from the embedding, which would then have the dual purpose of encoding lead time and time of day.

\begin{table}
    \centering
    \small
    \caption{
    Postprocessing models CRPS according to the conditioning strategy used for the lead time.
    The None column represents models that were not conditioned for the lead time.
    The Partition strategy uses a series of models trained on each time step individually.
    The Predictor strategy adds a lead time predictor. 
    The Embedding strategy injects a learned vector in the model input to represent the lead time.
}

\begin{tabular}{lcccc}
    \toprule
    & \multicolumn{4}{c}{Lead time conditioning strategy}   \\
    \cline{2-5}
    Model  & Partition  & Predictor  &  Embedding & Pred.+Emb. \\
    \midrule
    DRN   	& 1.655   &1.637&	1.638 &	\textbf{1.634} \\ 
    BQN    	& 1.641   &1.627&	1.626 &	\textbf{1.622}  \\
    QRN   	& 1.644   &1.638&	\textbf{1.634}	&1.635  \\
    \bottomrule
    \label{tab:condition}
\end{tabular}
\end{table}

\begin{figure*}
    \centering
    \includegraphics{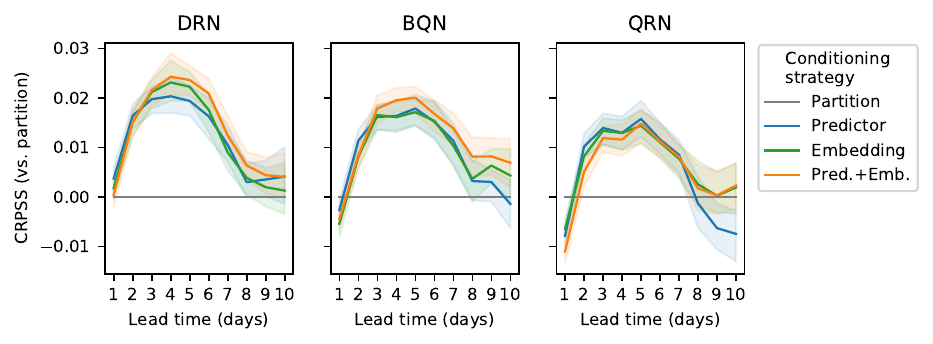}
    \caption{\gls{CRPSS} when training a NN for all lead times jointly.
    The baseline strategy is to train separate models for each lead time.
    The shaded areas represent 5 to 95\% confidence intervals.
    }
    \label{fig:step-condition}
\end{figure*}

Figure \ref{fig:step-condition} shows the performance of different conditioning strategies according to the lead time.
The confidence intervals are computed using the same bootstrapping strategy described in Section \ref{ssec:pp-performance}.
Interestingly, the benefits of training a single model vary by lead time.
The improvements are mostly observed around central lead times.
Early and late lead times perform similarly or worse than lead-time specific \glspl{NN}.
We envisage two explanations.
The first is related to the statistics of the data: the \gls{NN} converges to solutions that are well suited to intermediate uncertainties because they are the ``mean'' case in the dataset.
Another explanation can be considered which is related to predictability.
Since it is very high in early lead times, the postprocessing must be adapted to rely heavily on the forecast, which could benefit specialized models.
On the contrary, since predictability is very low at late lead times, it is difficult to do much better than a climatological model.
The benefits of training postprocessing models jointly would thus be concentrated in the intermediate lead times.

\begin{figure}
    \centering
    \includegraphics[width=0.8\columnwidth]{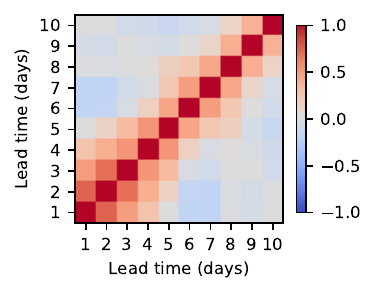}
    \caption{
    Lead time embedding self-similarity. 
    Every row and column corresponds to a learned vector representing a given lead time.
    The cells represent the degree of similarity between two of these vectors, as measured by the cosine similarity.
    This embedding was obtained by training a \gls{DRN} model.
    }
    \label{fig:self-similarity}
\end{figure}

We study the embedding learned to represent lead time in Figure \ref{fig:self-similarity}.
Our NN models learns 10 vectors $\mathbf{v}_i$, each representing a lead time.
Their mutual similarity is computed using the cosine similarity $s$, such that
\begin{align}
    s(\vec{v}_i, \vec{v}_j) &= \frac{\vec{v}_i \vec{v}_j}{\lVert \vec{v}_i \rVert \lVert \vec{v}_j \rVert}.
\end{align}
The learned embedding vectors have a diagonal similarity matrix, whereas nearby lead times are more similar.
Early lead times show more similarity to one another than later ones.
We suggest this is due to a shift in postprocessing strategies where early forecasts are strongly related to the underlying \gls{NWP} model, while later forecasts are more concerned with uncertainty quantification and knowledge of statistical trends in the data.

\subsection*{Benefits of ensemble forecast for probabilistic postprocessing}

The results from the previous subsections show that much of the uncertainty related to the forecast can be recovered statistically with a neural network.
A follow-up question is to what extent ensemble members are useful in estimating the forecast distribution, given that it is postprocessed.
To investigate this, we run an experiment on the ENS10 dataset~\citep{ENS10Dataset} where we progressively add ensemble members to the postprocessing model input.
ENS10 is a 10-member reforecast dataset built from outputs of an operational configuration of the ECMWF IFS model (cycles Cy43r1 and Cy45r1).
It spans a period going from January 1998 to December 2017, making two forecasts a week.
We postprocessed this \gls{NWP} model for lead times of one and two days.
More detailed information about the dataset is contained in Table \ref{tab:ens10-dataset}.

\begin{figure*}
    \centering
    \includegraphics[]{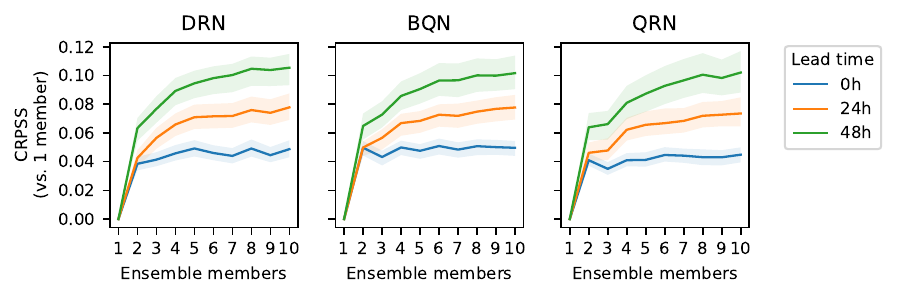}
    \caption{
    \gls{CRPSS} gain related to adding ensemble members to the postprocessing model input. 
    The dataset is ENS10.
    The baseline is training postprocessing using only the control member.
    The shaded areas represent 5 to 95\% confidence intervals.
    }
    \label{fig:benefits-of-ensemble}
\end{figure*}

To input the \gls{NWP}-dependent features from multiple ensemble members simultaneously, we applied the linear layer to each member individually, then averaged the resulting vectors.
For the \gls{DRN} model, the initial estimate $\hat{\sigma}$ was set to the standard deviation of the predicted value in the ensemble forecast when more than one member was available.
There exist more elaborate ways of leveraging multiple ensemble members for postprocessing~\citep{RaspNeuralNetworks2018,FinnSelfAttentiveEnsemble2021,HohleinPostprocessingEnsemble2024}.
Despite this, Figure \ref{fig:benefits-of-ensemble} shows how the forecast improves when adding ensemble members.
The metric is the \gls{CRPSS} computed against postprocessing only the control member.
The confidence intervals are computed using paired bootstrapping against the 1-member model.
Our work is mostly concerned with the left side of the figure where there is a sharp increase in performance where the second member is added, at all lead times.
We conclude that multiple ensemble members are indeed useful in making a probabilistic forecast, even in the presence of probabilistic postprocessing.
This conclusion is in line with previous work~\citep{BremnesEvaluationForecasts2023}.

We observe that the impact of supplementary ensemble members increases with lead time.
Furthermore, the impact of ensemble members seems to be diminishing.
However, it is difficult to conclude because our experimental setup is limited to a 48-hour lead time.
The importance of large ensembles is expected to increase with lead time, but this cannot be verified given the data available for this study.

\section{Discussion and conclusions}
\label{sec:discussion}

This work studied probabilistic forecasts produced from a deterministic \gls{NWP} model.
It evaluated the deterministic-to-probabilistic strategy under parametric and quantile-based assumptions with eight combinations of predictive models and uncertainty representations.
The best postprocessing models have a \gls{CRPSS} of about 15\% when compared against a naive probabilistic forecast and significantly outperform \gls{EMOS} for all but the latest lead times.
Under extreme conditions, the \gls{NN} model performance decreases in a way that is expected from a statistical model.

The proposed methodology can be applied to any numerical prediction model for which a sufficient training dataset is available.
We believe it has applications in domains that must rely on deterministic models but benefit from probabilistic decision-making, e.g. energy consumption forecasting.

To account for the large number of lead times to be encoded within one model, we introduced a lead time embedding.
This embedding brought modest improvements to the model \gls{CRPS}.
The learned representation has an intuitive post hoc interpretation.
Training a postprocessing model jointly for all lead times improved performance overall.
The improvements were concentrated around central lead times.

Our results with the NN show low sensitivity to the choice of uncertainty representation in terms of aggregated metrics.
However, their calibration exhibits noticeable differences, especially at the tails of the distribution.
This suggests the choice of an uncertainty representation should be evaluated with regard to its capacity to represent the full distribution.

While our approach is inexpensive, it still has caveats when compared to a postprocessed ensemble forecast.
Our experiment on the ENS10 dataset shows that adding ensemble members quickly brings important improvements to the \gls{CRPS}, indicating that supplementary numerical simulations are a robust way to improve the marginal forecast at a station.
Perhaps more importantly, our postprocessing strategy makes the independence assumption over all stations and lead times.
This breaks spatiotemporal consistency because it makes it impossible to sample from all stations and lead times in a way that is physically realizable~\citep{SchulzMachineLearning2022}.
Such consistency is an important asset when considering extreme events in downstream applications of the weather forecast.
This has been addressed notably with Schaake Shuffle~\citep{ClarkSchaakeShuffle2004,ShresthaUsingSchaake2020} and Ensemble Copula Coupling~\citep{SchefzikUncertaintyQuantification2013,LakatosComparisonMultivariate2023}, but these methods assume that the correlation structure between variables can be recovered using historical observations or the \gls{NWP} ensemble members, respectively.
This could be impossible in the presence of unresolved local effects.
As such, we identify avenues for future work inside and outside the independence assumption.

\subsection{Station and lead-time independent postprocessing}

For station and lead time-independent forecasting, we identify two directions to extend our work.
Firstly, the benefits of training postprocessing models jointly for all lead times could be investigated further.
Our experiments showed a tendency for \glspl{NN} postprocessing models to only make improvements in central lead times.
We suggested two interpretations of this phenomenon.
The first is related to dataset statistics where the model performs best in central lead times because they are at the center of the training distribution.
The second was related to the predictability of the weather itself where postprocessing strategies are different enough in early lead times and late lead times that specialized models do better.
The latter interpretation makes intuitive sense: one adjusts are 1-day forecast differently than a 10-day forecast to account for predictability.
Further experiments could be designed to identify which interpretation is correct.
As far as encoding the lead time is concerned, our experiments with the lead time embedding are moderately conclusive.
They did bring modest improvements to the \gls{CRPSS}, but not always in a statistically significant way.
The \gls{NN} models did not react in the same way to their introduction, with the \gls{DRN} benefitting more from it than the others.
This warrants further experimentations, notably where forecast validity time changes with lead time.
An embedding could let a neural network build a representation that efficiently blends the effects of lead time and time of day on postprocessing.

Secondly, our short experiment on the ENS10 dataset showed that the CRPS is quickly and decisively improved by adding ensemble members at the input.
This shows that supplementary numerical simulations are a robust way to improve the forecast.
The improvement may become even larger with \gls{NN} models whose architecture specifically leverages spread information from the ensemble.
Our experiment was also limited in lead time, and it would be of interest to perform it on more operational models at longer time horizons.
Since predictability declines with lead time, we expect the trend where later lead times benefit from more ensemble members to continue.
Further experiments are required to determine its shape on larger time horizons.

\subsection{Generative modeling for postprocessing}

Other lines of research are available outside of the station and lead time independence assumptions.
We believe that the generative modeling literature could help preserve spatiotemporal consistency for postprocessing at stations.
In computer vision, generative neural networks have been successful in sampling consistently from large output spaces.
Work has already been performed to that effect in postprocessing on grid~\citep{DaiSpatiallyCoherent2021} and in situ~\citep{ChenGenerativeMachine2024}, showing that spatial correlations can be recovered.
We expect similar methods could be used to represent temporal dependencies as well.
Questions remain about how to implement this exactly for in situ postprocessing.
We expect well-adapted architecture will propose a way to encode the spatial relationship between stations, as well as a generative component that is not subject to mode collapse and training instability concerns, which are common in generative modeling.

\acknowledgments
    This work was funded in part by Environment and Climate Change Canada, the Computer Research Institute of Montreal, and a Choose France Chair in AI grant from the French government. 
    Experiments were carried out using HPC resources from GENCI-IDRIS (Grant AD011014334).

\datastatement
    The operational archives used in this study are accessible on demand through the open data access program of the Meteorological Service of Canada~\citep{GDPSDataset}.
    The METAR observations used are freely available~\cite{METARDataset}.
    The source code of the models used in this work are available at \url{https://github.com/davidlandry93/pp2023/}.

\bibliographystyle{ametsocV6}
\balance
\bibliography{references_pp2023}

\pagebreak
\onecolumn
\appendix

\begin{table}[h]
\small
\centering
    \caption{ENS10 dataset}
    \begin{tabular}{cc}
    \toprule
    Dataset & ENS10 \\
    \midrule
    Time coverage & 1998 to 2017   \\
    Test period & 2015 to 2017 \\
    Frequency & Twice weekly  \\
    Initialization hour & 0000 UTC \\
    Lead times & 0h, 24h, 48h \\
    Resolution & Originally $\sim$36km, interpolated to 0.5\textdegree \\
    Ensemble Members & 10  \\
    NWP-dependent predictors & 88 \\
    \bottomrule
    \end{tabular}
\label{tab:ens10-dataset}
\end{table}
\begin{table*}[h]

    \centering
    \caption{NWP-dependent and NWP-independent predictors used in the ENS10 dataset. 
    1000..10 hPa denotes vertical levels 1000, 925, 850, 700, 500, 400, 300, 200, 100, 50 and 10 hPa. 
    \checkmark = Always used.}

    \small
    \begin{tabular}{llc}
    \toprule
    &Predictor                   &  ENS10  \\
    \midrule
    NWP-Dependent &Convective precipitation    & Column \\
    &Divergence                  & 1000..10 hPa \\
    &Geopotential                & 1000..10 hPa  \\
    &Mean sea-level pressure     & Sea-level datum \\
    &Skin temperature            & Surface \\
    &Specific humidity           & 1000..10 hPa  \\
    &Temperature                 & 2 m, 1000..10 hPa  \\
    &Total cloud cover           & Column \\
    &Total column water          & Column \\
    &Total column vertically-integrated water vapor & Column  \\
    &{U,V} component of wind     & 10 m, 1000..10 hPa \\
    &Wind speed                  & 10 m  \\
    \midrule
    NWP-independent&Lead time & \checkmark \\
    &Forecast day-of-year (sin and cos)  & \checkmark \\
    &Forecast time-of-day & \checkmark  \\
    &Latitude & \checkmark \\
    &Longitude& \checkmark \\
    &Elevation& \checkmark \\
    \bottomrule
    \end{tabular}
    \label{tab:ens10-features}
\end{table*}
\begin{figure*}
    \centering
    \includegraphics[]{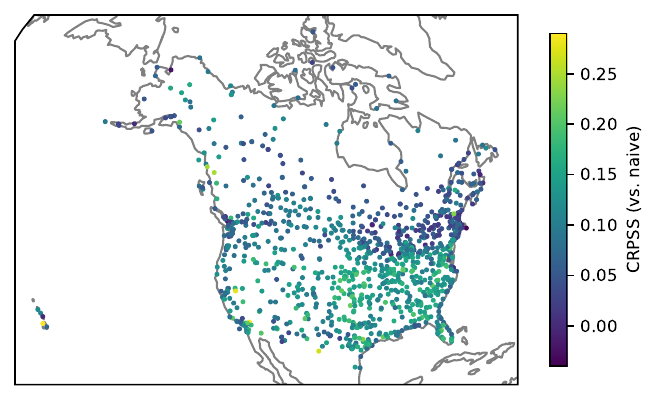}
    \includegraphics[]{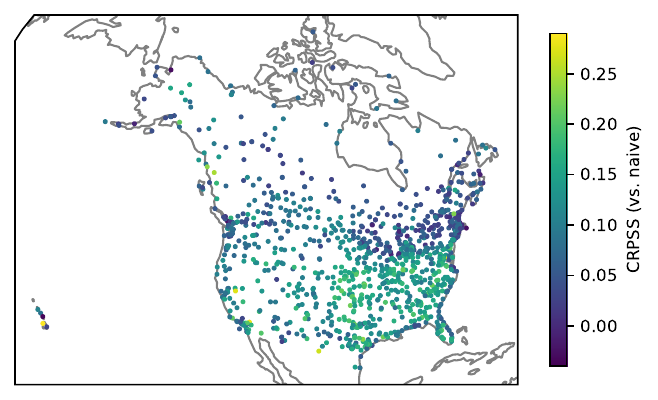}
    \caption{
    Skill gain brought by postprocessing models against a debiased baseline.
    \textbf{Top:} \gls{BQN} model.
    \textbf{Bottom:} \gls{QRN} model.
    }
    \label{fig:appendix-skill-gain-spatial}
\end{figure*}

\end{document}